\documentclass[12pt]{iopart}

\usepackage{iopams}  
\usepackage{color}

\begin{document}

\title[]{New Dirac Delta function based methods with applications to perturbative expansions in quantum field theory}

\author{Achim Kempf$^1$, David M. Jackson$^2$,  Alejandro H. Morales$^3$}

\address{$ $ \\ $^1$Departments of Applied Mathematics and Physics\\$^2$Department of Combinatorics and Optimization\\
University of Waterloo, Ontario N2L 3G1, Canada,\\ 
$^3$Laboratoire de Combinatoire et d'Informatique Math\'ematique (LaCIM)\\
Universit\'e du Qu\'ebec \`a Montr\'eal, Canada}

\begin{abstract}
We derive new all-purpose methods that involve the Dirac Delta distribution. Some of the new methods use derivatives in the argument of the Dirac Delta. We highlight potential avenues for applications to quantum field theory and we also exhibit a connection to the problem of blurring/deblurring in signal processing. We find that blurring, which can be thought of as a result of multi-path evolution, is, in Euclidean quantum field theory without spontaneous symmetry breaking, the strong coupling dual of the usual small coupling expansion in terms of the sum over Feynman graphs.   
\end{abstract}

\maketitle

\section{A method for generating new representations of the Dirac Delta}
The Dirac Delta distribution, see {\it e.g., \rm} \cite{sakurai,arfken,enc}, serves as a useful tool from physics to engineering. Our aim here is to develop new all-purpose methods involving the Dirac Delta distribution and to show possible avenues for applications, in particular, to quantum field theory. We begin by fixing the conventions for the Fourier transform:
\begin{equation}
\widetilde{g}(y) := \frac{1}{\sqrt{2\pi}} \int g(x) ~e^{ixy} ~dx \label{one}, \mbox{~~~}
g(x) = \frac{1}{\sqrt{2\pi}}\int \widetilde{g}(y)~ e^{-ixy}~ dy 
\end{equation}
To simplify the notation we denote integration over the real line by the absence of integration delimiters. 
Assume now that $g$ is a suitably well-behaved function. Then, 
\begin{eqnarray}
\delta(x) & = & \frac{1}{2\pi} \int e^{ixy}~dy \label{norm}\\
  & = & \frac{1}{2\pi} \int \frac{1}{g(y)}~g(y)~e^{ixy}~dy \\
  & = & \frac{1}{\sqrt{2\pi}}~\frac{1}{g(-i\partial_x)}~\frac{1}{\sqrt{2\pi}}\int g(y)~e^{ixy}~dy
\end{eqnarray}
and therefore:
\begin{equation}
\delta(x) = \frac{1}{\sqrt{2\pi}}~\frac{1}{g(-i\partial_x)}~\widetilde{g}(x) \label{reps1}
\end{equation}
Here, $g$ must be sufficiently well-behaved so that, in particular, $1/g$ has an expansion as a power series, which then gives meaning to $1/g(-i\partial_x)$ as a series in derivatives. 

By this method, each suitable choice of $g$ in Equ.\ref{reps1}  yields an exact representation of the Dirac Delta. 
For example, if we choose $g$ to be a Gaussian 
\begin{equation} 
g(x) := e^{-\sigma x^2/2}, ~~~~~~~~\widetilde{g}(x) = \frac{1}{\sqrt{\sigma}} ~e^{-x^2/2\sigma}
\end{equation} 
with $\sigma >0$, then:
\begin{eqnarray}
\delta(x) &=& \frac{1}{\sqrt{2\pi \sigma}}~e^{\frac{-\sigma}{2}\partial_x^2}~e^{-x^2/2\sigma}
\label{uu}
\\ &=&  \frac{1}{\sqrt{2\pi \sigma}}~
\sum_{n=0}^\infty \frac{ ({\frac{-\sigma}{2}\partial_x^2})^n}{n!} ~e^{-x^2/2\sigma} \label{sd}
\end{eqnarray}
We obtain approximations of $\delta(x)$ by truncating the series of derivatives in Equ.\ref{sd}. For example, when truncating the series after the first term, we obtain the standard Gaussian approximation $\delta(x) \approx (2\pi\sigma)^{-1/2} e^{-x^2/2\sigma}$ which converges to $\delta(x)$ as $\sigma \rightarrow 0$ in the weak limit.

\section{A representation of the Fourier transform through the Dirac Delta}
From Equ.\ref{reps1} we obtain a representation of the Fourier transform:
\begin{equation}
\label{fou1}
\widetilde{g}(x) =  {\sqrt{2\pi}}~{g(-i\partial_x)}~\delta(x) \label{fourier1}
\end{equation}
Let us test this representation of the Fourier transform by applying it to a basis of functions, namely the plane waves $g(x)=e^{ixw}$. Their Fourier transform must come out to be $\widetilde{g}(y)=\sqrt{2\pi}\delta(y+w)$, {\it i.e., \rm} $\widetilde{g}(x)=\sqrt{2\pi}\delta(x+w)$. We have $g(-i\partial_x)=e^{w\partial_x}$. Therefore, Equ.\ref{fourier1} indeed holds:
\begin{equation}
\sqrt{2\pi}~\delta(x+w) = \sqrt{2\pi}~e^{w\partial_x}\delta(x) 
\end{equation}
Also, we can use the fact that $g(x):=\delta(x)$ implies $\widetilde{g}(x)=  1/\sqrt{2\pi}$ to obtain from Equ.\ref{fourier1}:
\begin{equation}
\delta(-i\partial_x)\delta(x) = \frac{1}{2\pi} \label{deldel}
\end{equation}
While Equ.\ref{fourier1} is an exact representation of the Fourier transform, we obtain potentially useful approximations to the Fourier transform by replacing the Dirac Delta in Equ.\ref{fourier1} by one of its approximations. The derivatives in the power series expansion of $g(-i\partial_x)$ can then be carried out successively. For example, within the path integral formalism of quantum field theory (QFT), quantization is a Fourier transformation. In the case of a real scalar field, the generating functional $Z[J]$ of Feynman graphs (which we will here call `partition function' even in the non-Euclidean case) reads, see {\it e.g., \rm} \cite{weinberg}:
\begin{equation}
Z[J] = \int e^{iS[\phi]+i\int J \phi~d^nx}D[\phi] \label{pathintegral}
\end{equation}
$Z[J]$ is, therefore, the Fourier transform of $e^{iS[\phi]}$. Using Equ.\ref{fourier1}, we obtain 
\begin{equation}
Z[J] = N e^{iS[-i\delta/\delta J]} ~\delta[J] \label{partition1}
\end{equation}
where $N$ is a formally infinite normalization constant. Equ.\ref{partition1} invites approximating the partition function $Z[J]$ by using any one of the approximations for the Dirac Delta, such as a sinc, Lorentzian or Gaussian approximation on the right hand side (RHS). This then allows one to evaluate the RHS of Equ.\ref{partition1} explicitly, presumably best with the help of diagrammatics, similar to the usual use of Feynman rules. 
Recall that the usual pertubative expansion of $Z[J]$ is not convergent and is at best asymptotic. 
Here, the freedom to choose a regularization of $\delta[J]$ should translate into some freedom to affect the convergence properties of the diagrammatics. We will return to this issue. 
\section{A second representation of Fourier transformation using the Dirac Delta}
The new representation of the Fourier transform, Equ.\ref{fourier1}, features derivatives in the argument of the function that is to be Fourier transformed, $g(-i\partial_x)$, which may be inconvenient in some circumstances. We now give a representation of the Fourier transform which has the derivatives in the Dirac Delta instead:
\begin{equation}
\widetilde{g}(y) = \sqrt{2\pi}~e^{ixy}~\delta(i\partial_x -y)~g(x) \label{fourier2}
\end{equation}
In spite of appearances, the right hand side does not depend on $x$. To prove Equ.\ref{fourier2}, we first use Equ.\ref{one} to Fourier transform back and forth and then, in order to obtain the third line, we  integrate iteratively by parts with respect to $z$:
\begin{eqnarray}
& & \sqrt{2\pi}e^{ixy}\delta(i\partial_x -y)g(x)\nonumber \\
&  & =\frac{1}{2\pi}\int\int e^{izw}e^{-iwx}\sqrt{2\pi}e^{izy}\delta(i\partial_z-y)g(z) dzdw \nonumber \\
 &  & = \frac{1}{\sqrt{2\pi}}\int\int g(z) \delta(-i\partial_z-y) e^{iz(w+y)-iwx} dz dw \nonumber \\
 &  & =\frac{1}{\sqrt{2\pi}}\int\int g(z)\delta(w+y-y) e^{izy} e^{-iwx} dzdw \nonumber\\
 &  & =\frac{1}{\sqrt{2\pi}}\int g(z) e^{izy}  dz = \widetilde{g}(y)
\end{eqnarray}
Notice that we used $f(\partial_u)e^{\alpha u} = f(\alpha) e^{\alpha u}$ in the step from the third to the fourth line.
Let us test the representation of the Fourier transformation given by Equ.\ref{fourier2} by applying it to the plane waves basis, $g(x) = e^{ixw}$. We have to show that Equ.\ref{fourier2} yields $\widetilde{g}(y) =\sqrt{2\pi}\delta(y+w)$.  Indeed:
\begin{eqnarray}
\sqrt{2\pi} e^{ixy}\delta(i\partial_x-y) e^{ixw} &= &  \sqrt{2\pi} e^{ixy} \delta(-w-y) e^{ixw} \nonumber \\
 & = & \sqrt{2\pi} \delta(y+w)
\end{eqnarray}
Applied to quantum field theory, Equ.\ref{fourier2} yields:
\begin{equation}
Z[J] = N e^{i\int \phi J d^nx} ~\delta(i\delta/\delta\phi-J) e^{iS[\phi]} \label{partition2}
\end{equation}
Any approximation of the Dirac Delta which possesses a Taylor series expansion about the origin can be used in Equ.\ref{partition2} to yield another class of Feynman-like diagrammatic methods for evaluating $Z[J]$. 

\section{Integration in terms of the Dirac Delta}
Let us now use our two new representations of the Fourier transform to obtain two new representations of integration. To this end, we use the fact that integration is the zero-frequency case of the Fourier transform: 
$
\int g(x)~dx = \sqrt{2\pi}~\widetilde{g}(0).~
$
From our first representation of the Fourier transform, Equ.\ref{fourier1}, we therefore obtain:
\begin{equation}
\int g(x)~dx = 2\pi ~g(-i\partial_x)~\delta(x)|_{x=0} \label{int1}
\end{equation}
From our second representation of the Fourier transform, Equ.\ref{fourier2}, we obtain another representation of integration:
\begin{equation}
\int g(x)~dx = 2\pi~ \delta(i\partial_x)~g(x) \label{int2}
\end{equation}
We can also see this directly, using Equ.\ref{norm}:
\begin{equation}
2\pi \delta(i\partial_x)g(x) = \int e^{-w\partial_x} g(x)~dw=\int g(x-w)~dw = \int g(w)~dw
\end{equation} 
Applied to the quantum field theoretic path integral, Equ.\ref{pathintegral}, the equations Equ.\ref{int1} and Equ.\ref{int2} yield:
\begin{equation}
Z[J] = N e^{iS[-i\delta/\delta \phi]+\int J \delta/\delta\phi~d^nx}~\delta[\phi]|_{\phi=0}\label{partition3}
\end{equation}
\begin{equation}
Z[J] = N \delta[i\delta/\delta\phi] e^{iS[\phi]+i\int J \phi~d^nx} \label{partition4}
\end{equation}
Here, $N$ is a formally infinite normalization. 
Note that in Equ.\ref{partition4} the dependence on $\phi$ drops out in the same way that $x$ drops out in Equ.\ref{fourier2} and Equ.\ref{int2}. 

{
\section{Explicit examples, and applications to the Laplace transform}
\label{secfive}
In this section, we show that the new methods indeed add to the toolbox of practical techniques. To this end, we give examples for the Fourier transform and for integration over the real line, over bounded intervals and over semi-bounded intervals. As an example of the latter, we obtain a new method for evaluating the Laplace transform.
\subsection{Fourier transform}
For a simple application of the new Fourier transform method given in Equ.\ref{fou1}, let us calculate the Fourier transform $\widetilde{f}$ of $f(x)=\cos(x) e^{-x^2}$:
\begin{eqnarray}
 \widetilde{f}(x)  & = & \frac{1}{\sqrt{2\pi}}\int \cos(x)\,e^{-y^2}\,e^{i yx} dy\\
 &  & ~~~~\mbox{now using Equ.\ref{fou1}}: \\
& = & \sqrt{2\pi} \cos(-i\partial_x)~e^{-(-i\partial_x)^2}~\delta(x)\\
 & = & \sqrt{\frac{\pi}{2}} \left( e^{\partial_x}+e^{-\partial_x}\right)~e^{\partial_x^2}~\delta(x)\\
 &  & ~~~~\mbox{now expressing $\delta(x)$ through Equ.\ref{uu} with $\sigma =2$:}\nonumber \\
 & = & \frac{1}{2\sqrt{2}}~\left( e^{\partial_x}+e^{-\partial_x}\right)~e^{-x^2/4}\\
 & = & \frac{1}{2\sqrt{2}}~\left(e^{-(x+1)^2/4}+e^{-(x-1)^2/4}\right) \\
 & = & \frac{1}{\sqrt{2}e^{1/4}}e^{-x^2/4}\cosh(x/2)
\end{eqnarray}
\subsection{Integration}
Using Equ.\ref{int1}, we can evaluate straightforwardly, for example, the following integral:
\begin{eqnarray}
\int \frac{\sin(x)}{x}~ dx & = & 2\pi \frac{1}{2i}\left(e^{\partial_x}-e^{-\partial_x}\right)\frac{1}{-i\partial_x}~\delta(x)\vert_{x=0}\label{31}\\
 & = & \pi \left(e^{\partial_x}-e^{-\partial_x}\right) \left(\Theta(x) +c\right)\vert_{x=0}\\
 & = & \pi \left(\Theta(x+1) +c -\Theta(x-1)-c\right)\vert_{x=0}\label{33}\\
 & = & \pi
\end{eqnarray} 
Here, $\Theta$ is the Heaviside function and $c$ is an integration constant. 
Similarly, one readily obtains, e.g., $\int \sin(x)^5/x ~dx =
3\pi/8$,~ $\int \sin(x)^2/x^2~dx = \pi$, ~ $\int(1-\cos(tx))/x^2 dx=\pi \vert t\vert$ or $\int x^2 \cos(x) e^{-x^2} dx =\sqrt{\pi} e^{-1/4}/4$. 

Recall that the relationship between Equ.\ref{fou1} and Equ.\ref{int1} expresses integration as the zero frequency case of the Fourier transformation. Notice that this implies that 
when any integral is performed in the above way, i.e., through Equ.\ref{int1}, then the Fourier transform of the integrand is also immediately obtained, simply by dividing by $\sqrt{2\pi}$ and by not setting $x=0$.  For example, Eqs.\ref{31}-\ref{33} immediately also yield for $f(x)=\sin(x)/x$ that $\widetilde{f}(x) = \sqrt{\pi/2}\left(\Theta(x+1)-\Theta(x-1)\right)$. 


We remark that while the above integrals may also be integrated by other means, these tend to require a non-trivial search, such as a search for a suitable contour or for a suitable auxiliary integrating factor. The evaluation of the integrals with the new methods is more straightforward in the sense that the new methods allow one to evaluate the integrals essentially through the more direct task of the taking of derivatives. 

{	
\subsection{Integrals over semi-bounded intervals}
Our integration methods can also be applied to integrals over semi-bounded intervals. We begin with integrals over the interval $[0,\infty)$ and then extend this to the interval $[a,\infty)$ where a is real.  From Equ.\ref{int1}, we have: 
\begin{eqnarray}
\int_0^\infty f(x) \,dx &=&  \int_{-\infty}^\infty \Theta(x)f(x) dx = 2\pi \left. f(-i \partial_x) \Theta(-i \partial_x) \delta(x) \right|_{x=0} 
 \end{eqnarray}
But, from Equ.\ref{fou1}: 
\begin{equation}
\Theta(-i \partial_x) \delta(x) = \frac{1}{\sqrt{2\pi}} \widetilde{\Theta}(x).
\end{equation}
Moreover,
\begin{equation}
\widetilde{\Theta}(x) = \sqrt{\frac{\pi}{2}} \, \delta(x) +  \frac{1}{\sqrt{2\pi}} \, \mathsf{PP} \frac{i}{x}
\end{equation}
where $\mathsf{PP}$ denotes the ``principal value of\,".  Combining these three, we conclude that:
\begin{equation}\label{e:IntFormPP1}
\int_0^\infty f(x) \,dx = \left. f(-i \partial_x)  \left(\pi\, \delta(x) + \mathsf{PP} \frac{i}{x}\right)\right|_{x=0}.
\end{equation}
For integration over the interval $[a,\infty)$, where $a$ is real, this then also yields:
\begin{eqnarray}
\int_a^\infty f(x) dx = \int_0^\infty f(x+a) dx = \left.f(-i\partial_x +a) \left(\pi\, \delta(x) + \mathsf{PP} \frac{i}{x}\right)\right|_{x=0}.
\end{eqnarray}
The following example of an application of Equ.\ref{e:IntFormPP1} will now show that the Dirac Delta needs to be defined also in the upper complex plane. Assume $a$ is real. Then:
\begin{eqnarray}
\int_0^\infty e^{-a x} dx &=& e^{-ia\partial_x}\left.\left(\pi\delta(x) + \mathsf{PP}\frac{i}{x}\right)\right|_{x=0}\\
&=& \pi \left(\delta(ia)+\frac{1}{a}\right)
\end{eqnarray}
Comparing with the known result $\pi/a$ for Re$(a)>0$ and `undefined' for Re$(a)<0$, we conclude that the definition of the Dirac Delta needs to be extended to the upper half of the complex plane: $\delta(x)=0$ when Im$(x)>0$, while $\delta(x)$ has to remain undefined when Im$(x)<0$. 
The definition of $\delta(x)$ on the real line remains unchanged, of course. 
\subsection{Laplace transforms}
We now apply the above results to the Laplace transform. Equ.\ref{e:IntFormPP1} yields, for $a>0$, the following new representation of the Laplace transform:
\begin{eqnarray}
\label{e:LaTr:1}
\int_0^\infty e^{-a x}f(x) dx
 & = & \left.f(-i\partial_x) e^{ia \partial_x}
\left( \pi \delta(x) + \mathsf{PP} \frac{i}{x}\right)\right|_{x=0} \\
& = & \left.f(-i\partial_x) 
\left( \pi \delta(x+ia) + \mathsf{PP} \frac{i}{x+ia}\right)\right|_{x=0}\label{lp}
\end{eqnarray}
Let us test the new representation of the Laplace transform given in Equ.\ref{lp} by applying it to the basis of monomial functions $x^n$, where $n$ is any non-negative integer:  
\begin{eqnarray*}
\int_0^\infty e^{-a x} x^n dx
&=& \left. \left(-i\partial_x\right)^n \left( \pi \delta(x+ia) + \mathsf{PP} \frac{i}{x+ia}\right)\right|_{x=0} \\
&=& \left.\pi(-i)^n \delta^{(n)}(x+ia)\right|_{x=0}
+ \left.  i(-i)^n (-1)^n n! (x+ ia)^{-n-1}\right|_{x=0} 
\end{eqnarray*}
where $\delta^{(n)}(x) := \partial_x^n \delta(x).$  Recalling that $\delta(x)$ as well as its derivatives vanish when Im$(x)>0$, yields indeed:
\begin{equation}
\int_0^\infty e^{-a x} x^n dx
=  n! \, a^{-n-1} 
\end{equation}
We now use Equ.\ref{lp} to evaluate an example of a more non-trivial Laplace transform: 
\begin{eqnarray*}
\int_0^\infty \frac{\sin(wx)}{x} e^{-ax}dx
&=&  \frac{1}{2i} \left( e^{iw(-i\partial_x)} - e^{-iw(-i\partial_x)}\right) \frac{1}{-i\partial_x}\\
 & & \left. \left(\pi\delta(x+ia) +\frac{i}{x+ia}\right)\right|_{x=0} \\
&=& \left.  \frac{i}{2} \left( e^{w\partial_x} - e^{w(-\partial_x)}\right)\left(c+ \log(x+ia)\right) \right|_{x=0} \\
&=& \frac{i}{2} \left( \log(ia+w) - \log(ia-w) \right) \\
&=& \frac{i}{2} \log \left(\frac{ia+w}{ia-w}\right) = \tan^{-1}\left(\frac{w}{a}\right).
\end{eqnarray*}
Notice that, in the second line, $\delta(x+ia)=0$ because $a>0$. 
\subsection{Integrals over a bounded interval}
Finally, let us consider examples of integrals over a finite interval $[a,b]$ of the real line. To this end, we notice that, from Equ.\ref{fou1} and Equ.\ref{int1}, integrals over a function $f$, weighted by a function $g$, can be written in the following form:
\begin{eqnarray}
\int f(x) g(x) dx & = & 2\pi\, f(-i\partial_x) g(-i\partial_x)\delta(x)\vert_{x=0}\\
 & = & \sqrt{2\pi}\, f(-i\partial_x) \widetilde{g}(x)\vert_{x=0}\label{wf}
\end{eqnarray}
Let us choose for $g$ the characteristic function of the interval $[a,b]$
 \begin{equation}
 g(x) :=
\left\{
\begin{array}{ccc}
1  &  \mbox{if $x\in[a,b]$,}    \\
0  &    \mbox{if $x\not\in[a,b]$}
\end{array}
\right.
\end{equation}
which has the Fourier transform: 
\begin{equation}
\widetilde{g}(x) = \frac{1}{\sqrt{2\pi}\, i x} \left( e^{ibx} - e^{iax}\right). \label{e:fWy}
\end{equation}
Thus, Equ.\ref{wf} yields this new representation of integration over a finite interval:
\begin{equation}\label{e:abInt}
\int_a^b f(x) dx = \left.  f(-i \partial_x) \frac{1}{ix} \left( e^{ibx} - e^{iax}\right) \right|_{x=0}
\end{equation}
Let us test it by applying it to the basis of monomial functions $x^n$. Indeed, Equ.\ref{e:abInt} yields:
\begin{eqnarray*}
\int_a^b x^n dx 
&=&  \left. -i\,  (-i \partial_x)^n \frac{1}{x} \left( e^{ibx} - e^{iax}\right) \right|_{x=0}  \\
&=& -i (-i\partial_x)^n \sum_{k=1}^{n+1}\left. \left(\frac{(ib)^k x^{k-1}}{k!}-\frac{(ia)^k x^{k-1}}{k!}\right)\right|_{x=0}\\
&=&  \frac{b^{n+1} - a^{n+1}}{n+1}.
\end{eqnarray*}
Notice that, for example, Fourier series coefficients $\int_a^b x^n e^{ixy}dx$ then also immediately follow. Notice also that a change of variables in Equ.\ref{e:abInt} yields for the anti-derivative: 
\begin{equation}
\int^x f(x') dx' = f(\partial_y)(e^{xy}-1)/y|_{y=0}+c
\end{equation}   
\section{Blurring and deblurring using the Dirac Delta}
\label{secsix}
For a concrete application of the new methods let us now apply one of our formulas above, Equ.\ref{reps1}, in the context of an important problem in scientific \cite{puetteretal} and engineering signal processing \cite{image}, namely to the problem of deblurring signals such as images. The process of blurring is the mapping of a signal, $f$, into a blurred signal, $f_B$, through a convolution. In one dimension:
\begin{equation}
f_B(y):=\int f(x)~g(x-y)~dx
\end{equation}
For example, in the case of so-called Gaussian blurring, the blurring kernel reads:
\begin{equation}
g_{Gauss}(x) := \sqrt{\frac{a}{2\pi}}~e^{-\frac{a}{2}x^2} \label{g-blur}
\end{equation}
While analytic Fourier methods for deblurring are well known in the signal processing community, we can use the above results to describe generic deblurring in a new and very compact way. Namely, our claim is that if $g$ is such that the  blurring by $g$ is invertible, then the deblurring can be implemented through an operator $D_g$ which can be written as $D_g := 1/(\sqrt{2\pi} \widetilde{g}(i\partial_y))$, {\it i.e.: \rm}
\begin{equation}
D_g~f_B(y) = \frac{1}{\sqrt{2\pi}\widetilde{g}(i\partial_y)}~f_B(y) = f(y) \label{deblur}
\end{equation}
Here, we assume that $1/\widetilde{g}$ possesses a power series expansion about the origin. 
To prove Equ.\ref{deblur}, we use Equ.\ref{reps1}:
\begin{eqnarray}
\frac{1}{\sqrt{2\pi}\widetilde{g}(i\partial_y)}~f_B(y) & = & \int f(x)~\frac{1}{\sqrt{2\pi}\tilde{g}(i\partial_y)}~g(x-y)~dx\nonumber \\ \nonumber
 &  & (\mbox{use:}~z:=x-y, dx=dz,\partial_y=-\partial_z)\\ \nonumber 
 & = & \int f(y+z)~\frac{1}{\sqrt{2\pi}\widetilde{g}(-i\partial_z)}~g(z)~dz\\
 & = & \int f(y+z)~\delta(z)~dz = f(y)
\end{eqnarray}
For example, in the case of the Gaussian blurring kernel of Equ.\ref{g-blur}, we readily find our deblurring operator:
\begin{equation}
D_{Gauss} = e^{-\frac{1}{2a}\partial_y^2} = \sum_{n=0}^\infty \frac{1}{n!}\left(\frac{-\partial_y^2}{2a}\right)^n \label{degauss}
\end{equation}
Let us recall that the exponentiation of a single derivative, $e^{a\partial_x}$, has a simple interpretation as a translation operator, $e^{a\partial_x}f(x)=f(x+a)$. 
Equ.\ref{degauss} shows that the operator obtained by exponentiating the second derivative possesses an interpretation as a deblurring operator. 
Let us consider applying the series expansion on the RHS of Equ.\ref{degauss} term by term on a blurred function $f_B(y)$. In this way, we obtain a series expansion of the deblurring operation.  Intuitively, a blurred function is one in which fine details have been suppressed. The lowest order term, $1$, reproduces the blurred image. The higher order terms add derivatives of the function with suitable positive and negative prefactors. The derivatives enhance fine details of the function. Eventually, by adding more and more terms of the series of derivates in Equ.\ref{degauss}, the original image is restored. 

While this picture is accurate, it must be used with caution because simply adding derivatives would not suffice to sharpen an image. The particular coefficients and, in particular, the minus sign in Equ.\ref{degauss} are providing crucial cancellations in the series. To see this, notice that the inverse of the deblurring operator, $D^{-1}_{Gauss}$, which is the same as $D_{Gauss}$ except for the minus sign, is blurring rather than deblurring. 

The blurring operator also provides intuition for our integration formula Equ.\ref{int2}. Namely, 
using the usual Gaussian regularisation of the Dirac Delta, Equ.\ref{int2} can be written in this form:
\begin{equation}
\int g(x)~dx = 2\pi~\lim_{\epsilon\rightarrow 0} \frac{1}{\sqrt{2\pi\epsilon}}~e^{\frac{1}{\epsilon}\partial_x^2}~g(x) 
\end{equation}
We now see that, as $\epsilon\rightarrow 0$,  the area under the function $g$ flows apart in the sense that 
$g$ becomes more and more blurred. The amplitudes of $g$ would eventually drop to zero, were it not for the prefactor $1/\sqrt{\epsilon}$ that compensates so that the resulting amplitude is the original area. 
} 
Indeed, the special case of Gaussian blurring is also the case of the heat equation with the blurring operator being the heat kernel. Heat kernel methods, see {\it e.g., \rm} \cite{Davies}, then also link up with spectral methods, see {\it e.g., \rm} \cite{AK}.   

\section{A connection between deblurring and Euclidean quantum field theory}
We will now show that Euclidean quantum field theoretical path integrals naturally contain Gaussian deblurring operators. These deblurring operators possess a series expansion which then yields a series expansion of the partition function $Z[J]$. To see this, let us consider the example of Euclidean $\phi^4$ theory with $m^2>0$ and $\lambda>0$, i.e., without spontaneous symmetry breaking:
\begin{equation}
Z[J] = \int e^{\int d^nx-\frac{1}{2}\phi(x)(-\Delta+m^2)\phi(x) -\lambda \phi^4(x) + J(x)\phi(x)}D[\phi] \label{Z}
 \end{equation}
The deblurring operator originates in the kinetic term of the action: we replace the occurrences of $\phi(x)$ in the quadratic term in the action by $\delta/\delta J(x)$ and we move the exponentiated 
quadratic term in front of the path integral, to obtain: 
\begin{eqnarray}
Z[J] &=& e^{-\frac{1}{2}\int d^nx~\frac{\delta}{\delta J(x)}(-\Delta+m^2)\frac{\delta}{\delta J(x)}}\nonumber\\
 & & \times \int e^{\int d^nx~ -\lambda \phi^4(x) + J(x)\phi(x)}D[\phi]\label{newexp}
 \end{eqnarray}
Since the Laplacian, $-\Delta$, is a positive operator, the term before the path integral is a Gaussian deblurring operator and the resulting expansion is convergent. 
The deblurring operator is diagonal in the momentum representation and deblurs every mode $J(p)$ by a different amount, given by the scale $p^2+m^2$. Notice, that in the case of QFT on Minkowski space the extra imaginary unit in the exponent of the term before the path integral would not allow one to interpret this term as a deblurring operator. We will discuss the related issue of Wick rotation further below. What, then is the physical meaning of this deblurring expansion in Euclidean QFT? 
It is the strong coupling expansion which was first introduced in \cite{benderetal}. To see this, we change variables in Equ.\ref{newexp} from  $J(x_i)$ to $R(x_i):=\lambda^{-1/4} J(x_i)$ and carry out the remaining hypergeometric integrals to obtain
\begin{eqnarray}
Z[\lambda^{1/4}R] &=& c~ e^{-\frac{1}{2\sqrt{\lambda}} \int d^nx~ \frac{\delta}{\delta R(x)}(\Delta + m^2) \frac{\delta}{\delta R(x)}} \nonumber\\
& &~ \times ~ e^{\int d^nx' \log(P(R(x')))} \label{new}
\end{eqnarray}
where we recognize the expansion in powers of $1/\sqrt{\lambda}$. Here, $c$ is a constant that is independent of $R$ and 
\begin{eqnarray}
P(r)&=& \label{P}
\frac{1}{4\Gamma(\frac{3}{4})}\, \left(  \left( \Gamma  \left(
3/4
\right)  \right) ^{2}{r}^{2}
{{~{}_0F_2}\left(\frac{5}{4},\frac{3}{2},{\frac {1}{256}}\,{r}^{4}\right)}\right. \nonumber\\ 
& & \mbox{~~~~~~~~~~} \left. +2\,\sqrt {2}\pi \,
{{~{}_0F_2}\left(\frac{1}{2},\frac{3}{4},{\frac {1}{256}}\,{r}^{4}\right)} \right) 
\end{eqnarray}
where ${}_0F_2$ is a generalized hypergeometric function. $P(r)$ can easily be Taylor expanded, so that the perturbative expansion in $1/\sqrt{\lambda}$ in Equ.\ref{new} can be carried out.

In contrast, in order to derive the Feynman rules for the small coupling expansion, starting from Equ.\ref{Z}, one normally replaces occurrences of $\phi(x)$ in the interaction term by $\delta/\delta J(x)$ and then moves the interaction term in front of the path integral:
\begin{eqnarray}
Z[J] &=& e^{-\int d^nx~\lambda (\delta/\delta J(x))^4(x) } \\
 & & \times \int e^{-\frac{1}{2}\int d^nx~\phi(x)(-\Delta+m^2)\phi(x) +\int d^nx~ J(x)\phi(x)}D[\phi] \label{oldexp} \nonumber
 \end{eqnarray}
As is well known, see {\it e.g., \rm} \cite{klauder,nonanalyticity}, this step is analytically not justified. Indeed, since the Euclidean path integral is of the form 
\begin{equation}
z(j) = \int e^{-a x^2-\lambda x^4 +j x}dx
\end{equation}
any expansion in $\lambda$ about $\lambda=0$ must have a vanishing radius of convergence. This is because $z(j)$ diverges for all negative $\lambda$. As a consequence, Equ.\ref{oldexp} yields a perturbative expansion in $\lambda$ which is divergent. The reason why the small coupling expansion is nevertheless useful is of course that the expansion happens to be asymptotic, {\it i.e., \rm} it approaches the correct value to some extent before diverging.   

Finally, let us recall that the weak and the strong coupling expansions in Euclidean and Minkowski space QFT are obtained by expressing either the interaction terms or the kinetic terms of the action through functional differentiations respectively. Equ.22, which we derived in Minkowski space, opens up a new possibility, in which neither part of the action is expressed in terms of functional derivatives. Instead, a leading Dirac Delta is expressed in terms of functional derivatives. In principle, the Dirac Delta in Equ.22 can be replaced by any one of its representations, e.g., in terms of sequences of Gaussians or sinc functions. The resulting expansion of Z[J] should be neither a weak nor a strong coupling expansion and its properties should therefore be very interesting to determine. 


\section{Summary and Outlook}
We obtained new representations of the Dirac Delta in Equ.\ref{reps1}, of the Fourier transform in Equs.\ref{fourier1},\ref{fourier2}, of integration in Equs.\ref{int1},\ref{int2} and of the Laplace transform in Equ.\ref{lp}. Notice that, since our new methods are meant to serve as all-purpose methods, we intentionally did not use Wick rotation in their derivation (or anywhere else), because that would have required restriction to special cases in which poles are absent in certain regions of the complex plane. In particular, we did not use Wick rotation in our derivation of the new representation of the Laplace transform in Equ.\ref{lp}.  

The new all-purpose methods should find applications in a range of fields from physics to engineering. In particular, the new methods could be useful in practical calculations of Fourier transforms, Laplace transforms and integrals in general, as we showed in Sec.\ref{secfive} by giving explicit examples. An advantage of the new methods is that they do not involve a search for a suitable contour or a suitable integrating factor. Instead, the new methods essentially rely on differentiation, which is more straightforward. 

The new methods could also be useful in formal calculations, by enabling one to derive and represent relationships in a succinct new way. Further, the Dirac Deltas in the new representations of Fourier and Laplace transforms and of integration may sometimes be usefully replaced by approximations of the Dirac Delta. The new methods therefore offer numerous new ways in which integrals and Fourier and Laplace transforms, such as those occurring in QFT, can be regulated or approximated. 

In particular, with Equs.\ref{partition1},\ref{partition2},\ref{partition3},\ref{partition4} we obtained new ways to represent quantum field theoretical partition functions. More representations of partitions functions can be derived with the new methods, for example, by using the new representation of the Laplace transform, Equ.\ref{lp}. 
Notice that our new representations of partition functions offer a formal alternative to using path integration, since they involve only functional differentiations and the Dirac Delta. The new techniques invite application in formal calculations, such 
as derivations of Dyson Schwinger equations and Ward or Slavnov-Taylor identities. This then also leads, for example, to the question of how the effects of a nontrivial path integral measure will manifest themselves within the new integral-free approach. In particular, when the partition function is expressed as a path integral, anomalies arise from a nontrivial transformation behavior of the path integral's measure, as Fujukawa's method shows \cite{fujikawa}.  How do anomalies arise in the new representations of the partition function that do not involve path integration? 
Further, the new representations of the partition functions invite the exploration of the resulting perturbative expansions of the partition functions. This includes, in particular, also those expansions that arise when replacing the Dirac Delta by a regular function that approximates the Dirac Delta and that possesses a power series expansion. Of interest in this context are then, for example, the convergence properties and the physical interpretation of these expansions of $Z[J]$. 

In order to illustrate that the new methods possess concrete applications, } we then applied Equ.\ref{reps1} in Sec.\ref{secsix} to obtain a compact and transparent expression, namely Equ.\ref{deblur}, for a generic deblurring operator for signals. 
We also showed that deblurring operators naturally occur in Euclidean QFT. In particular, we identified the large coupling expansion as a deblurring expansion. Blurring, being a convolution, can be viewed as a sum over `paths'. In a sense, therefore, in quantum field theory the sum over paths that occurs in the phenomenon of (Gaussian) blurring is the strong coupling version of the small coupling sum over `paths' given by the the usual sum over Feynman graphs. 

A very interesting problem is the development of the functional analytic and distribution theoretic underpinnings of the  
new uses of the Dirac Delta, in particular, when the latter has derivatives in its argument, and also to study the algebraic and combinatorial ramifications for operations in rings of formal power series. For the necessary combinatorial background, see \it e.g., \rm \cite{jackson}. 
$$$$
\noindent \bf Acknowledgement: \rm AK and DMJ acknowledge support from NSERC's Discovery program. AHM was supported by a CRM-ISM postdoctoral fellowship. 
$$$$


\section*{References}
\begin{thebibliography}{11}

\bibitem{sakurai} J. J. Sakurai, J.J. Napolitano, \it Modern Quantum Mechanics, \rm second Ed., Addison-Wesley, Boston (2010)

\bibitem{arfken} G.B. Arfken, H.J. Weber, \it Mathematical Methods for Physicists (5th ed.), \rm Academic Press, Boston (2000) 

\bibitem{enc} K. Ito, S Nihon, \it Encyclopedic Dictionary of Mathematics, Second Ed. \rm MIT Press, Cambridge MA, USA (2000)

\bibitem{weinberg} S. Weinberg, \it The Quantum Theory of Fields II, \rm CUP, Cambridge, U.K., (1996)

\bibitem{puetteretal} R.C. Puetter, T.R. Gosnell, A. Yahil, \it Digital image reconstruction: Deblurring and Denoising, \rm 
Annual Review of Astronomy and Astrophysics, Vol. 43: 139-194 (2005)

\bibitem{image} B.K. Gunturk, X. Lin (Eds.), \it Image restoration: fundamentals and advances, \rm CRC Press, Boca Raton (2013)

\bibitem{Davies} E.B. Davies, \it Heat Kernels and Spectral Theory, \rm Cambridge University Press, Cambridge, U.K. (1990)

\bibitem{AK} A. Kempf, Phys. Rev. Lett. \bf 103, \rm 231301 (2009)

\bibitem{benderetal} C.M. Bender, F. Cooper, G.S. Guralnik, D.H. Sharp, Phys. Rev. \bf D19\rm, 1865 (1979) 

\bibitem{jackson} I.P. Goulden, D.M. Jackson, \it Combinatorial Enumeration, \rm Dover, Mineola N.Y. (2004)


\bibitem{klauder} J.R. Klauder, \it Beyond Conventional Quantization, \rm Cambridge University Press, Cambridge, U.K. (2000)
\bibitem{nonanalyticity}  R. Ticciati, \it Quantum Field Theory for Mathematicians, \rm Cambridge University Press, Cambridge U.K. (1999)

\bibitem{fujikawa} K. Fujikawa, Phys. Rev. Lett., \bf 42, \rm 1195 (1979)


\end{thebibliography}
\end{document}